\documentclass[sigconf]{acmart}

\def\BibTeX{{\rm B\kern-.05em{\sc i\kern-.025em b}\kern-.08emT\kern-.1667em\lower.7ex\hbox{E}\kern-.125emX}}
    
\usepackage{subcaption}

\copyrightyear{2019}
\acmYear{2019}
\setcopyright{acmlicensed}
\acmConference[FDG '19]{FDG '19: Procedural Content Generation Workshop}{August 26-30, 2019}{San Luis Obispo, CA}
\acmBooktitle{FDG '19: Procedural Conten Generation Workshop, August 26-30, 2019, San Luis Obispo, CA}
\acmPrice{15.00}
\acmDOI{10.1145/1122445.1122456}
\acmISBN{978-1-4503-9999-9/18/06}

\begin{document}

%
\title{Organic Building Generation in Minecraft}

%
\author{Michael Cerny Green}
\email{mcg520@nyu.edu}
\affiliation{%
  \institution{New York University}
}
\author{Christoph Salge}
\email{ChristophSalge@gmail.com}
\affiliation{%
  \institution{University of Hertfordshire}
}

\author{Julian Togelius}
\email{julian@togelius.com}
\affiliation{%
  \institution{New York University}
}

%
\begin{abstract}
This paper presents a method for generating floor plans for structures in Minecraft (Mojang 2009). Given a 3D space, it will auto-generate a building to fill that space using a combination of constrained growth and cellular automata. The result is a series of organic-looking buildings complete with rooms, windows, and doors connecting them. The method is applied to the Generative Design in Minecraft (GDMC) competition~\cite{salge2018generative} to auto-generate buildings in Minecraft, and the results are discussed.
\end{abstract}

%
%
%

%
\maketitle

\section{Introduction}
Procedural Content Generation in games~\cite{shaker2016procedural, togelius2011search} (PCG) comes in a variety of flavors. AI has been shown to excel in the automatic creation of levels~\cite{khalifa2016general, compton2006procedural}, narrative~\cite{rowe2009storyeval}, tutorials~\cite{green2018atdelfi, green2017press}, levels for tutorials \cite{green2018generating, khalifa2019intentional}, puzzles~\cite{khalifa2015automatic, shaker2013automatic}, and even entire games~\cite{khalifa2017general, green2018data}. The functionality and acceptable similarity of the content depends on the genre the AI generates for, but it is generally desired that high quality content can be generated rapidly on-demand. 

However, certain PCG algorithms are known to suffer from repetition, based on the nature of the algorihtms themselves. Rule-based generative agents are known to create good content that looks similar~\cite{shaker2016procedural}.  On the other hand, search-based agents can create diverse content but must spend time to ensure that this diverse content is functional for the player~\cite{togelius2011search}. The challenge then becomes a balance of similarity, time, functionality, and above all else, being able to paramaterize and customize the generator's output.

In this paper we apply a constrained growth method to Minecraft (Mojang 2009) within the Generative Design in Minecraft competition~\cite{salge2018generative} to create functionally similar but diverse looking floor plans for structures. This method is simple compared to others. Its strength lies in its speed, being able to generate a thousand fully connected structures in just a few seconds (Section \ref{sec:analysis_metrics}). A wide selection of structures generated by the algorithm are curated and are discussed below. Although constrained rectangular-growth applied to floor plan generation is not new, the novelty of our method lies in its modification to the algorithm to not require rectangular growth, instead using the natural Minecraft block units. The resulting structures work similarly to each other but look uniquely distinct. To our knowledge, this is also the first application of any PCG method in regards to full Minecraft structure generation.

We want to point out that all code discussed in this project is publicly available on the official GDMC Github page.\footnote{https://github.com/mcgreentn/GDMC} The code can easily be modularly applied to any existing settlement generator in the GDMC challenge in order to automatically generating structures.

\section{Background}
Building and structure generators are a well-explored area of PCG\cite{shaker2016procedural}, in part caused by the relatively recent increase of virtual worlds and environments that required large quantities of content. This section will discuss previous research in city and structure generation in a variety of domains, as well the Generative Design in Minecraft competition, the particular domain this paper explores.

\subsection{City and Structure Generation}
City, structure, and building generation is a popular PCG application, within and outside of games. Citigen~\cite{kelly2007citygen}, an automatic city generation system, is an example that generates the the urban geometry of a modern city. Given a terrain model, the system develops a system of road networks and building footprints, which can be used to place buildings manually or automatically. Kelly et~al.~\cite{kelly2006survey} survey a collection of different city generation techniques, including techniques such as geometric rule-based systems~\cite{greuter2003real, greuter2003undiscovered, greuter2004beyond}, L-systems~\cite{parish2001procedural}, agent-based simulation~\cite{lechner2003procedural}, template-based systems~\cite{sun2002template}, and split-grammars~\cite{wonka2003instant}.

The above methods describe ways of generating collections of structures or even entire cities, but there are also attempts to design the internal ``floor-plans'' of buildings. The \textit{LaHave House project.} by Rau-Chaplin
et al.~\cite{rau1996lahave} generates floor plans using shape grammars. Hahn et al.~\cite{hahn2006persistent} developed a system which generates office building rooms in a just-in-time way, based on player movement and positioning. Martin~\cite{martin2006procedural} researched a graph-based method, which treats rooms as nodes and connections between rooms as edges, with user-defined constraints like \textit{room count} and \textit{room function}. Tutenel et al.~\cite{tutenel2009rule} developed a rule-based system which defines room types using a semantic database, and entities can develop relationships between their adjacent neighbors. Lopes et al.~\cite{lopes2010constrained} tested an constrained rectangular L-growth algorithm which generated fully-connected rooms for structures, which heavily influenced the ideas in this project. Camozzato et al.~\cite{camozzato2015procedural} procedurally develop floor-plans using a hand-drawn building exterior as input in a rectangular-growth-based approach. Guo and Li~\cite{guo2017evolutionary} created a system which uses a combined approach of agent-based search and optimization to created multi-level structure floor-plans.

\subsection{Minecraft \& The Generative Design in Minecraft Competition}
Minecraft (Mojang 2009) is a popular open-world survival game where the player is spawned within a voxel-block world. Gameplay largely consists of ``mining'' blocks and building tools and structures with them. While the game has an ``developer designed'' goal for the player to accomplish- defeating the Ender Dragon - many players rather focus on building houses or settlements. Every coordinate location in Minecraft can be represented as a $1x1x1$ block. \textit{All blocks, positions, and dimensions mentioned in this paper are in regards to this representation}.

The Generative Design in Minecraft (GDMC) AI settlement generation competition~\cite{salge2018generative} is a new AI challenge in which the goal is to develop algorithms that can develop adaptive and ``interesting'' cities and towns in Minecraft. Instead of ``clean-slate'' generation done by many existing PCG systems, where the generator is not restricted by already-existing game elements, this competition focuses on ``adaptive'' generation, where the generator is required to build on top of and in response to artifacts that already exist. For example, if a river exists within the given map terrain, a generator would build a bridge across that river.

The GMDC has just closed submissions for its second iteration of the settlement generator competition, and one observation is that none of the entries for the last two rounds have attempted to fully generate structures during generation. Most structures are based on templates and are often varied using hard-coded variables like dimensionality and building material. The internal floor-plans of these structures are almost non-existent or also based on templates. This paper attempts to push the boundaries of this by fully generating the internals of structures from scratch. For this reason, we are making available all of the code used in this project, which we attempted to make as easy as possible to implement in any settlement generator by simplifying the input: \textit{given a rectangular 3 dimensional space, generate a floor-plan for it}.

\section{Methods}
The building generator described in this paper can be divided into two parts: 1.) floor plan generation and 2.) external wall generation. These methods are discussed below.

\subsection{Floor Plan Generation}
The floor plan is generated using a simple constrained growth algorithm. The original idea for using this as a floor plan generator stemmed off an L-shaped constrained growth method researched by Lopes et al.~\cite{lopes2010constrained}. Unlike their method, this one does not grow in a rectangular fashion but instead uses the natural granular unit provided in Minecraft, \textit{a single block}.

\subsubsection{Room Placement}
\begin{figure}
    \centering
    \begin{subfigure}[t]{.47\linewidth}
        \includegraphics[width=0.95\linewidth]{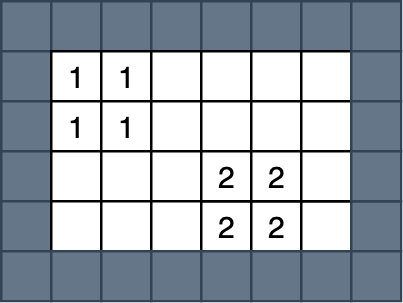}
        \caption{Initialized starting positions}
        \label{fig:growing_rooms_a}
    \end{subfigure}
    \centering
    \begin{subfigure}[t]{.47\linewidth}
        \includegraphics[width=0.95\linewidth]{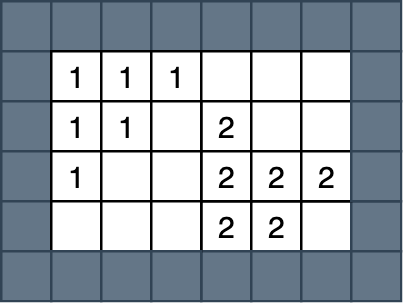}
        \caption{$2$ turns into growing}
        \label{fig:growing_rooms_b}
    \end{subfigure}
    \centering
    \begin{subfigure}[t]{.47\linewidth}
        \includegraphics[width=0.95\linewidth]{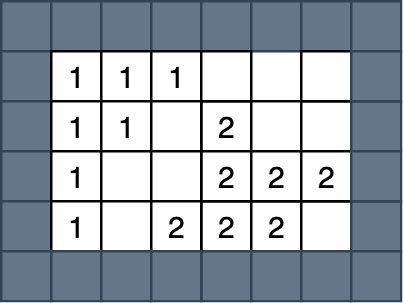}
        \caption{No more room $1$ growth}
        \label{fig:growing_rooms_c}
    \end{subfigure}
    \centering
    \begin{subfigure}[t]{.47\linewidth}
        \includegraphics[width=0.95\linewidth]{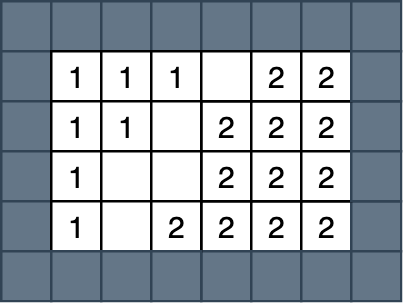}
        \caption{Finished growing}
        \label{fig:growing_rooms_d}
    \end{subfigure}
    \caption{A walkthrough of the constrained growth process}
    \label{fig:growing_rooms}
\end{figure}

First, the number of rooms is calculated by taking the rounded cubed root, i.e. $\sqrt[3]x$, of the total area of the rectangle in which the building will be generated. For example, obeying this rule, a structure that is $9x9$ will be 4 rooms. This does not take the height of the building into account. After the number of rooms is determined, each room is given a random initial starting location within the space. Rooms are not allowed to start on top of, or directly adjacent to each other. A room's starting location is designated by a $2x2$ square. Figure \ref{fig:growing_rooms_a} shows an example of initial room placement. All rooms must start within the external walls of the building (the darker blocks). If a room cannot be placed, e.g. the placement of previous rooms prevent it from finding applicable space after 100 attempts, the room is no longer considered a part of the building plans.

\subsubsection{Room Growth}
After initial room placement, the rooms take turns growing themselves by one block each turn, until none of the rooms are able to grow any more. The order in which this is done is randomized, as room turns are shuffled after each iteration through all the rooms. Figure \ref{fig:growing_rooms_b} displays the results of the constrained growth algorithm for the same building in Figure \ref{fig:growing_rooms_a} after two iterations. The rules of growth are simple. On its turn, a room searches for potential growth locations, determined by their direct adjacency (not diagonal adjacency) to other blocks already in the room. A block that is already adjacent to another room is not considered a potential growth candidate. Rooms also cannot grow into the external walls. Each room is growing in an organic way, with no motivation to retain its initial geometrically square shape. Figure \ref{fig:growing_rooms_d} displays the rooms after both have run out of room to grow.

\subsubsection{Door Placement}
\begin{figure}
    \centering
    \begin{subfigure}[t]{.47\linewidth}
        \includegraphics[width=0.95\linewidth]{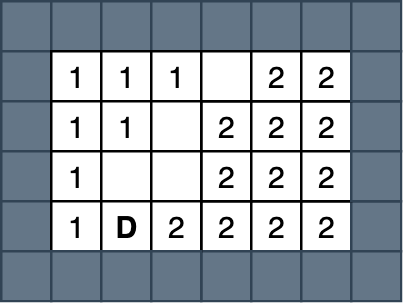}
        \caption{Door placement 1}
        \label{fig:door_placement_a}
    \end{subfigure}
    \centering
    \begin{subfigure}[t]{.47\linewidth}
        \includegraphics[width=0.95\linewidth]{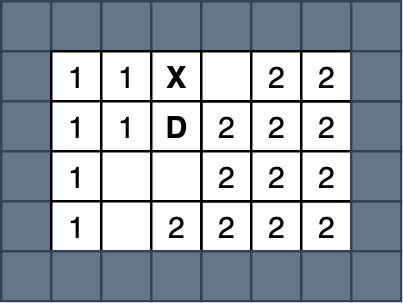}
        \caption{Door placement 2}
        \label{fig:door_placement_b}
    \end{subfigure}
    \centering
    \begin{subfigure}[t]{.47\linewidth}
        \includegraphics[width=0.95\linewidth]{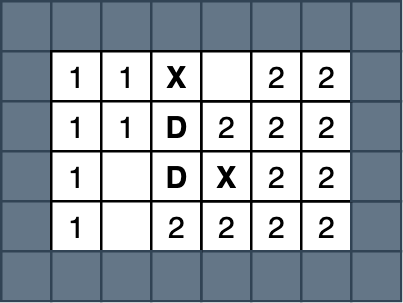}
        \caption{Door placement 3}
        \label{fig:door_placement_c}
    \end{subfigure}
    \centering
    \begin{subfigure}[t]{.47\linewidth}
        \includegraphics[width=0.95\linewidth]{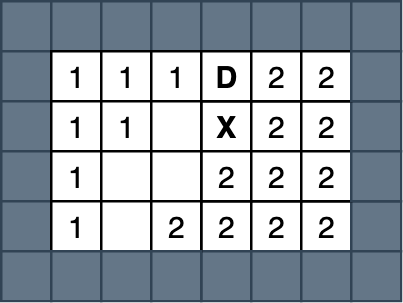}
        \caption{Door placement 4}
        \label{fig:door_placement_d}
    \end{subfigure}
    \caption{Various legal door placement possibilities. 'D' marks a door that has been placed. 'X' marks a tile once designated as a room area that has been transformed into a wall because of door placement.}
    \label{fig:door_placement}
\end{figure}

After rooms are finished growing, the generator moves onto  the door placement segment of the process. In this stage, doors are placed to connect rooms, and a single door is placed in the external wall to let one into the building. Unlike some generative methods which use an optimization graph-connectivity algorithm, door placement is done on a granular level. A door can only be placed where there is a wall and where it would be "in-between" to two different rooms (or another door). In addition, doors can only be placed if it is adjacent to at least one other wall, and when a door is placed, it will place two walls on either side of it. Figure \ref{fig:door_placement} displays $4$ different legal ways that doors might be placed in the previous building example. In some cases, several of these placement possibilities may occur simultaneously, so that rooms have multiple doors between them. Figure \ref{fig:growing_rooms_c} shows an example of two doors placed adjacently, which can create the in-game effect of either one room being extended by a block or even making a small hallway.

\subsection{External Wall Generation}
After the floor plan has been generated, the system begins the process of creating the external walls. This is done using a process that comes from a family of algorithms known as Cellular Automata (CA), a strategy well-known in games and simulations. In this system, CA is used to self-organize the placement of solid blocks and glass windows, to the effect of creating interesting exterior walls for structures.

This system uses a style of neighbor summation, or in other words, not caring so much about specific neighbor states, but the summation of those states. A block state is characterized as a $1$ if it is a window, and $0$ if it is a solid block. Each block has 4 neighbors and themselves; therefore sum of the states can lie anywhere between $0$ and $5$. At wall initialization, a matrix of the height at width/depth of the building (depending on which wall is being generated) is randomly generated with $75\%$ of the wall being solid blocks and $25\%$ being glass blocks. The rules for cellular automata are simple: if the sum equals $2$ or $3$, the current block is a glass block, otherwise it is a solid block. After $10$ generations, the wall is considered finished. After generating all $4$ walls, they are placed into the sides of the building. An external door into one of the rooms is also placed randomly in one of the external walls, as an entrance into the building.

\section{Analysis}
The data from our structural analysis comes from $3$ experiments in which structures of various dimensions were generated. Each experiment generated $1,000$ buildings. The first explored small $7x7$ block buildings (3-room), the second skinny $6x12$ block buildings (3-room), and the third larger $15x15$ block buildings (5-room). Several metrics are measured over the course of generating buildings and are described in Section \ref{sec:analysis_metrics}. In Section \ref{sec:analysis_observations}, we present a subjective evaluation of the general organic feel of the buildings that are generated and provide screenshots of a few generated artifacts in Minecraft.

\subsection{Metrics}\label{sec:analysis_metrics}
All experiments were done on a MacBook Pro 2016 with a $2.9$ GHz Intel Core i$5$ processor and $8$ GB of RAM.  Experiment 1 (small buildings) took approximately $7.90$ seconds, and Experiment 2 (long skinny buildings) took approximately $12.30$ seconds, Experiment 3 (large buildings) took approximately $154.03$ seconds. Experiments were conducted on a single thread. As expected, the larger the dimensions of the building become, the more time is required for the constrained growth and cellular automata algorithms to complete.

\begin{figure}
    \centering
    \begin{subfigure}[t]{1\linewidth}
        \includegraphics[width=0.95\linewidth]{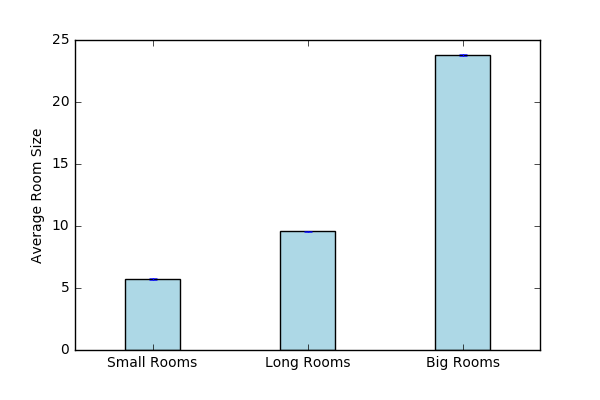}
        \caption{Average room area per building}
        \label{fig:metrics_a}
    \end{subfigure}
    \centering
    \begin{subfigure}[t]{1\linewidth}
        \includegraphics[width=0.95\linewidth]{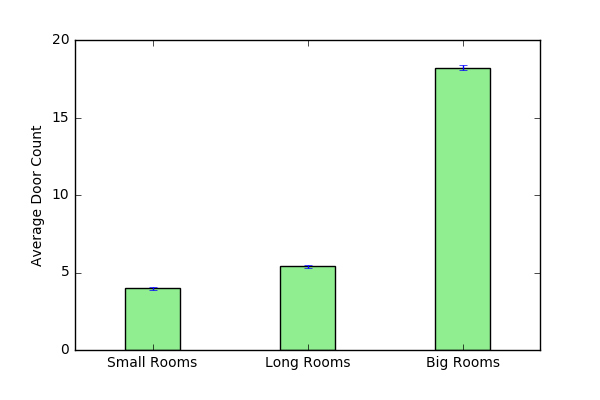}
        \caption{Average door count per building}
        \label{fig:metrics_b}
    \end{subfigure}
    \caption{Metrics collected during experimentation; all buildings are fully connected. Error bars represent true mean $95\%$ confidence intervals}
    \label{fig:metrics}
\end{figure}

In addition to time spent generating, $2$ building metrics are measured over the course of generation: number of doors and average room size. We want to point out that all buildings were fully connected (i.e. one could traverse all areas of empty space in the building). Figure \ref{fig:metrics} displays these metrics.

\begin{itemize}
    \item \textit{$7x7$ Small Buildings}: These structures all contain $3$ rooms each. The average room size is approximately $5.72$ with a 95\% confidence interval of $0.069$. There are  $3.92$ doors on average with a 95\% confidence interval of $0.072$.
    \item \textit{$6x12$ Long Buildings}: These structures also all contain $3$ rooms each. The average room size is approximately $9.56$ with a $95\%$ confidence interval of $0.069$ There are $5.40$ doors on average with a $95\%$ confidence interval of $0.081$
    \item \textit{$15x15$ Large Buildings}: These structures all contain $5$ rooms each. The average room size is approximately $23.78$ with a $95\%$ confidence interval of $0.061$ There are $18.21$ doors on average with a $95\%$ confidence interval of $0.180$
\end{itemize}

The most obvious conclusion one can draw from these numbers is that as the dimensions of the structures increase (and thus square block-age increases), there is an exponential increase in the average room area. This is because room-count is determined by taking the cubic root of raw square block-age. Another conclusion is that average door count per building also increases at a seemingly exponential rate. Even in a building with only 5 rooms, there is an average of roughly $18$ doors, suggesting that there are large inefficiencies in connectivity. However, this might make sense if, for example, the rooms in question were very long and might require multiple entrances and exits.

\subsection{Observations}\label{sec:analysis_observations}
In this section, we discuss a curated group of structures of various sizes and layouts. 

\begin{figure}
    \centering
    \begin{subfigure}[t]{0.47\linewidth}
        \includegraphics[width=0.95\linewidth, height=8.96cm]{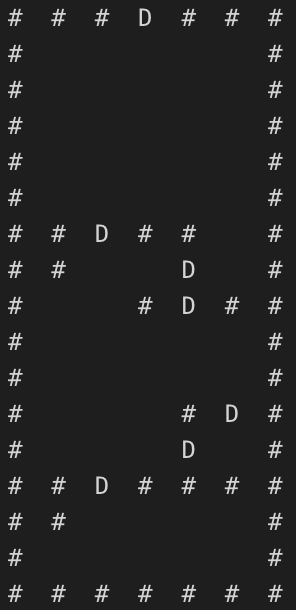}
        \caption{Generated ASCII layout}
        \label{fig:birdseye-layout_a}
    \end{subfigure}
    \centering
    \begin{subfigure}[t]{0.47\linewidth}
        \includegraphics[width=0.95\linewidth]{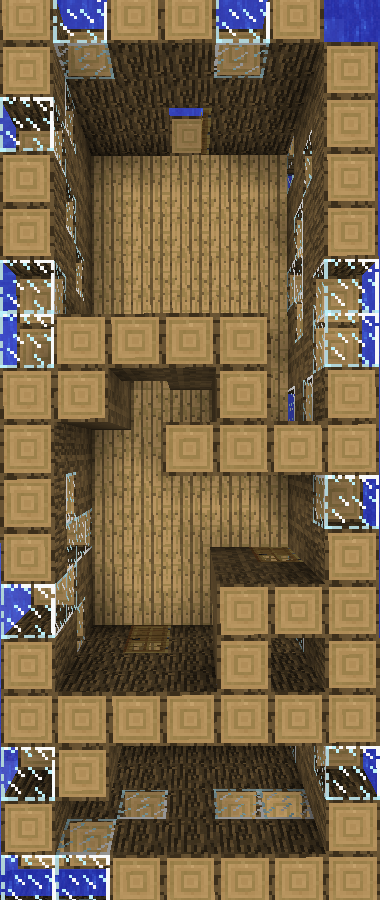}
        \caption{The generated building}
        \label{fig:birdseye-layout_b}
    \end{subfigure}
    \caption{A generated ASCII character building layout compared to a Minecraft building generated using that layout}
    \label{fig:birdseye-layout}
\end{figure}

\begin{figure}
    \centering
    \begin{subfigure}[t]{1\linewidth}
        \includegraphics[width=0.95\linewidth, height=6cm]{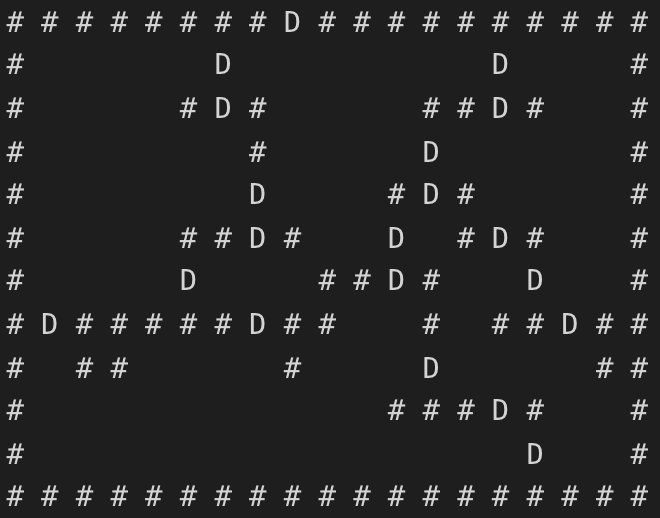}
    \end{subfigure}
    \centering
    \begin{subfigure}[t]{1\linewidth}
        \includegraphics[width=0.95\linewidth, height=6cm]{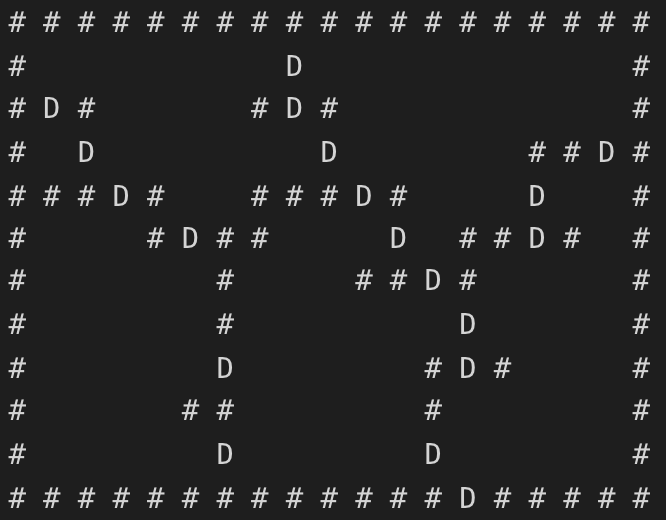}
    \end{subfigure}
        \begin{subfigure}[t]{1\linewidth}
        \includegraphics[width=0.95\linewidth, height=6cm]{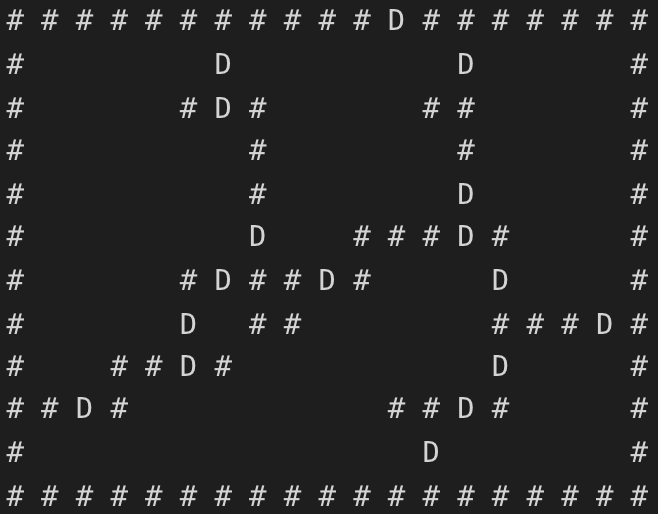}
    \end{subfigure}
    \caption{Examples generated layouts using identical dimensions (12 x 19)}
    \label{fig:examples}
\end{figure}

Figure \ref{fig:birdseye-layout} displays a generated ASCII building layout and the resulting building in Minecraft. Figures \ref{fig:examples} display several examples of generated floor-plans using the same dimensional space. Unlike real-world structures, the layout of these rooms are organic looking and much less rectangular by comparison. Small alcoves and closet areas are more commonplace, and one can see a variety of odd shapes. 

\begin{figure}
    \centering
    \includegraphics[width=\columnwidth]{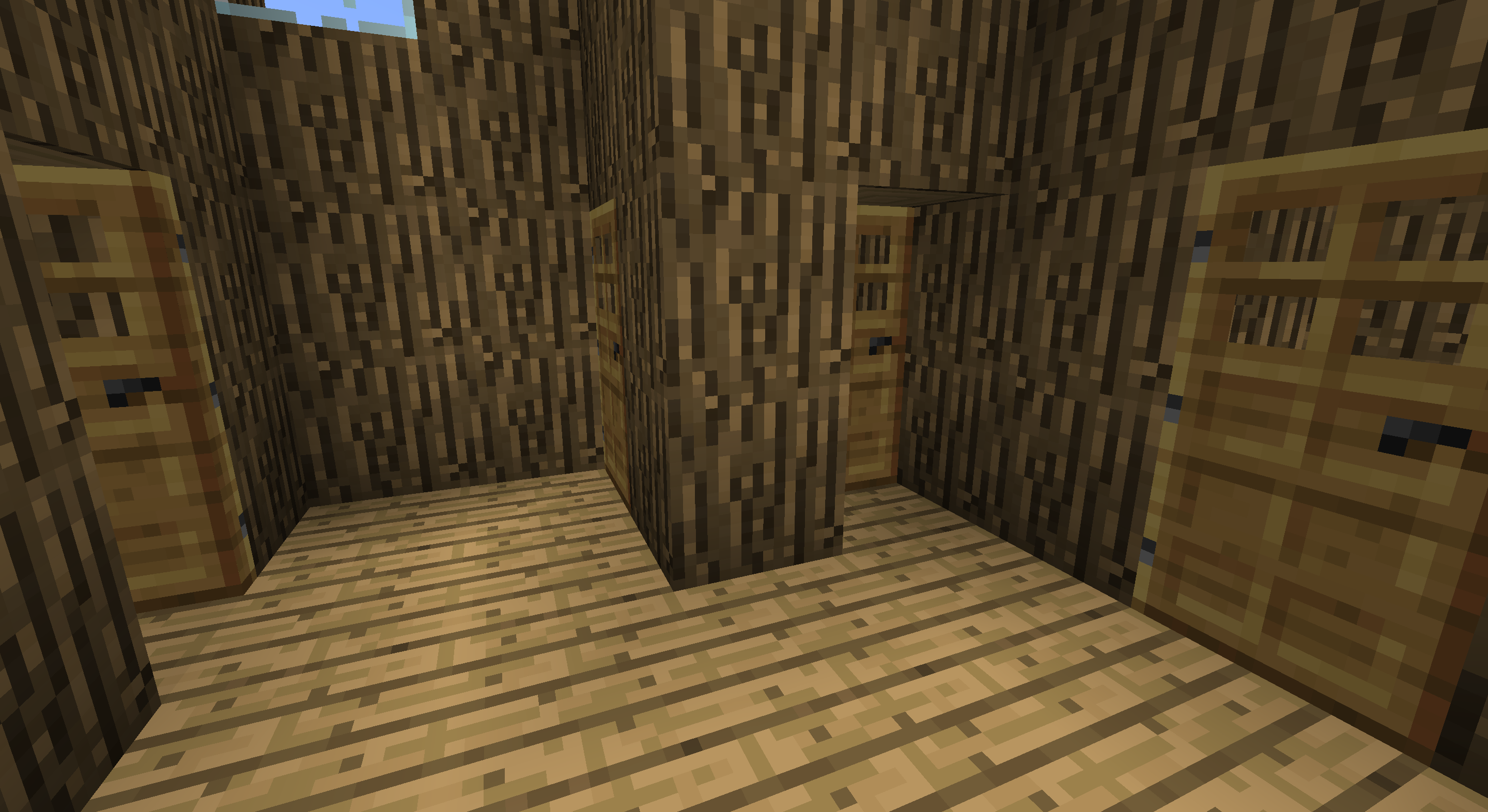}
    \caption{Large structures have larger rooms, which often contain many choices}
    \label{fig:choices}
\end{figure}

\begin{figure}
    \centering
    \includegraphics[width=\columnwidth]{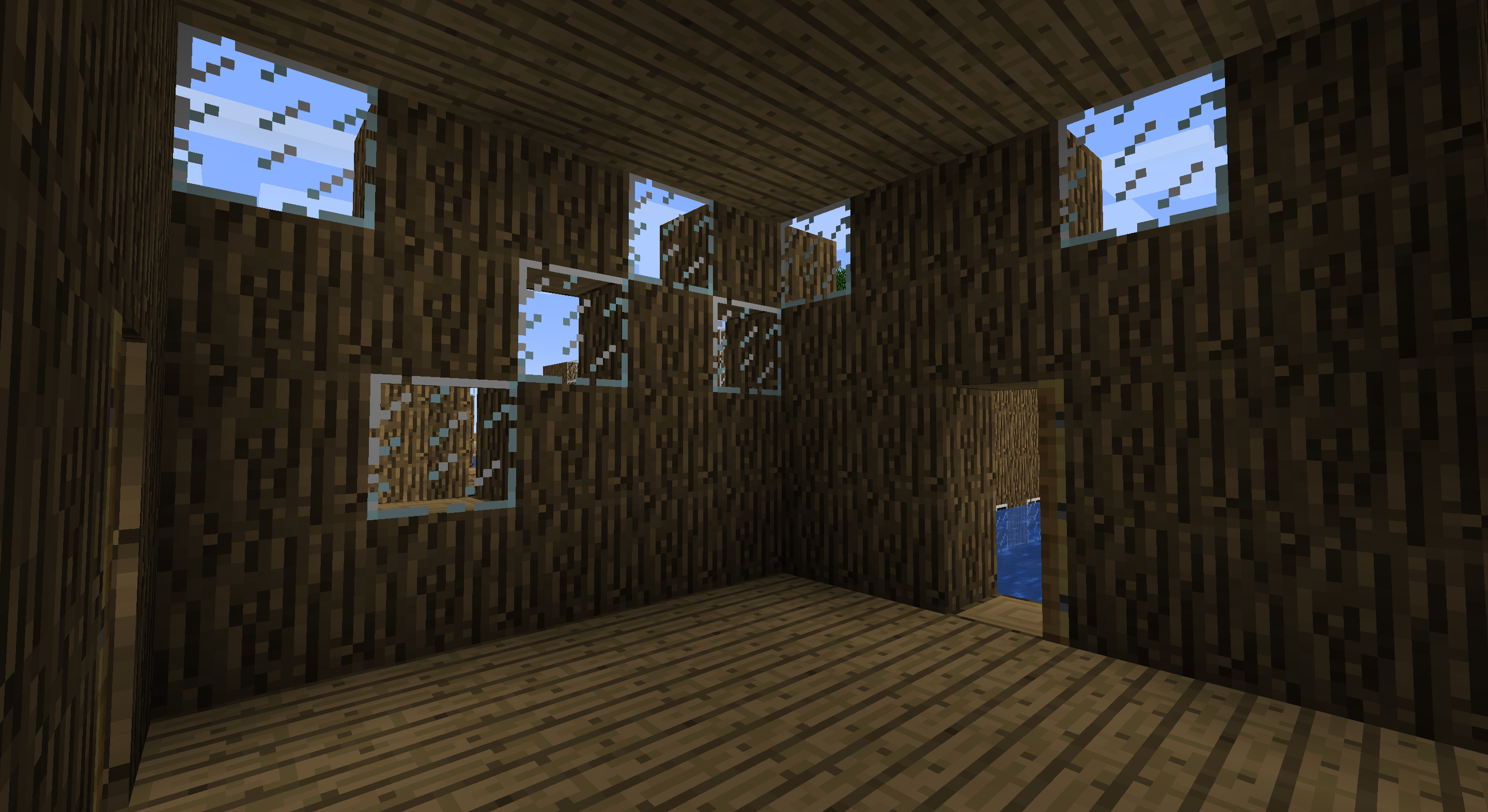}
    \caption{Caption}
    \label{fig:living-room}
\end{figure}

When the buildings are small, they are simple. The lower average door count (compared to the larger structures) means that the buildings themselves are more linear. Figure \ref{fig:living-room} displays a room that could defined as a living room or a bedroom. As structures get larger, there are many more pathways through the structure. Sometimes this can be disorienting without any identifying features in the house. In addition to this, large buildings often have bigger rooms, and therefore more doors per room in close proximity. Figure \ref{fig:choices} displays such an example of a large amount of door choices. In the future, a generator which includes an interior decoration system would be of great benefit in preventing confusion.

\begin{figure}
    \centering
    \includegraphics[width=\columnwidth]{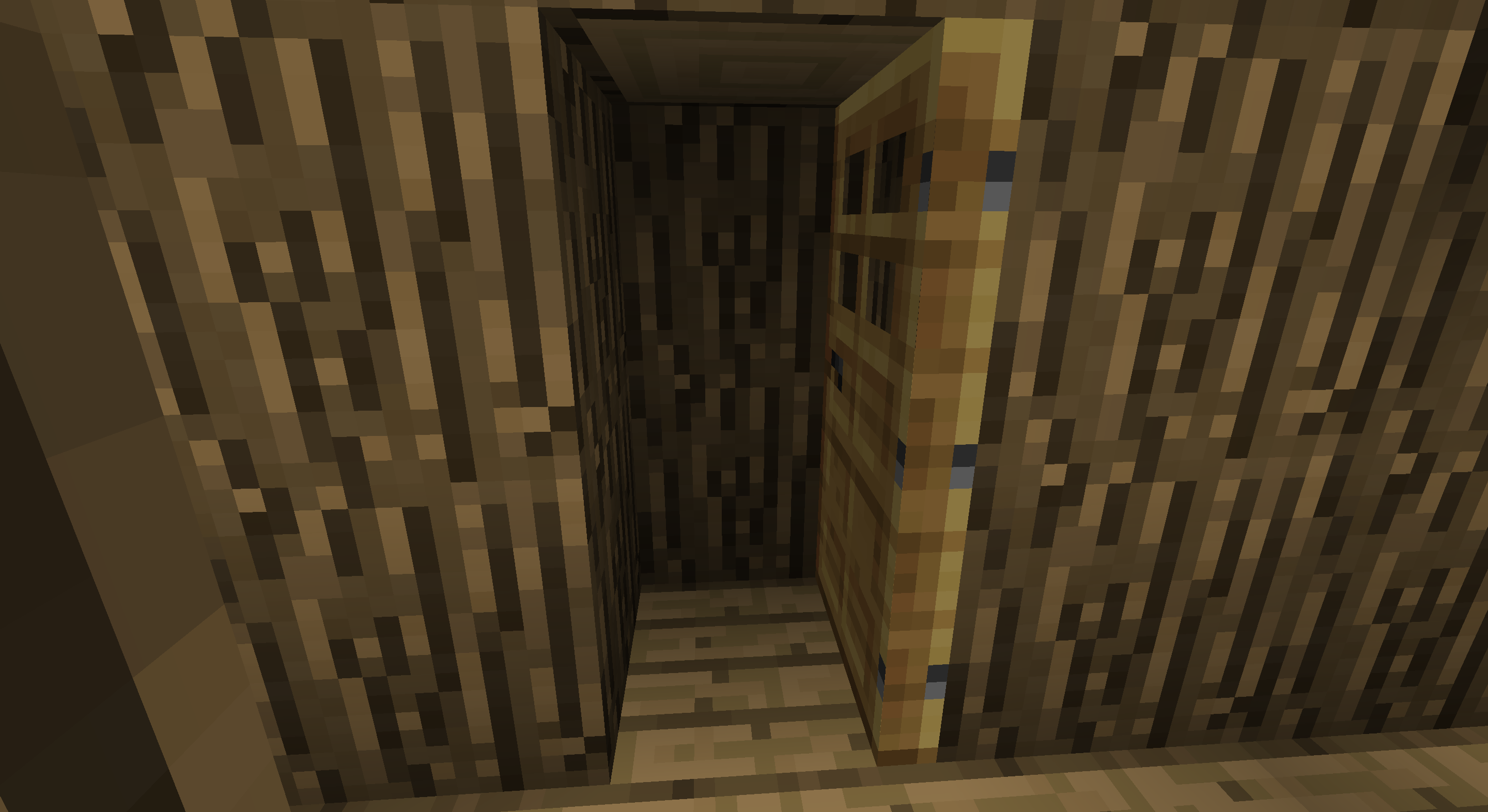}
    \caption{Looking into a small room generated as a side effect of door placement}
    \label{fig:small-room1}
\end{figure}

\begin{figure}
    \centering
    \includegraphics[width=\columnwidth]{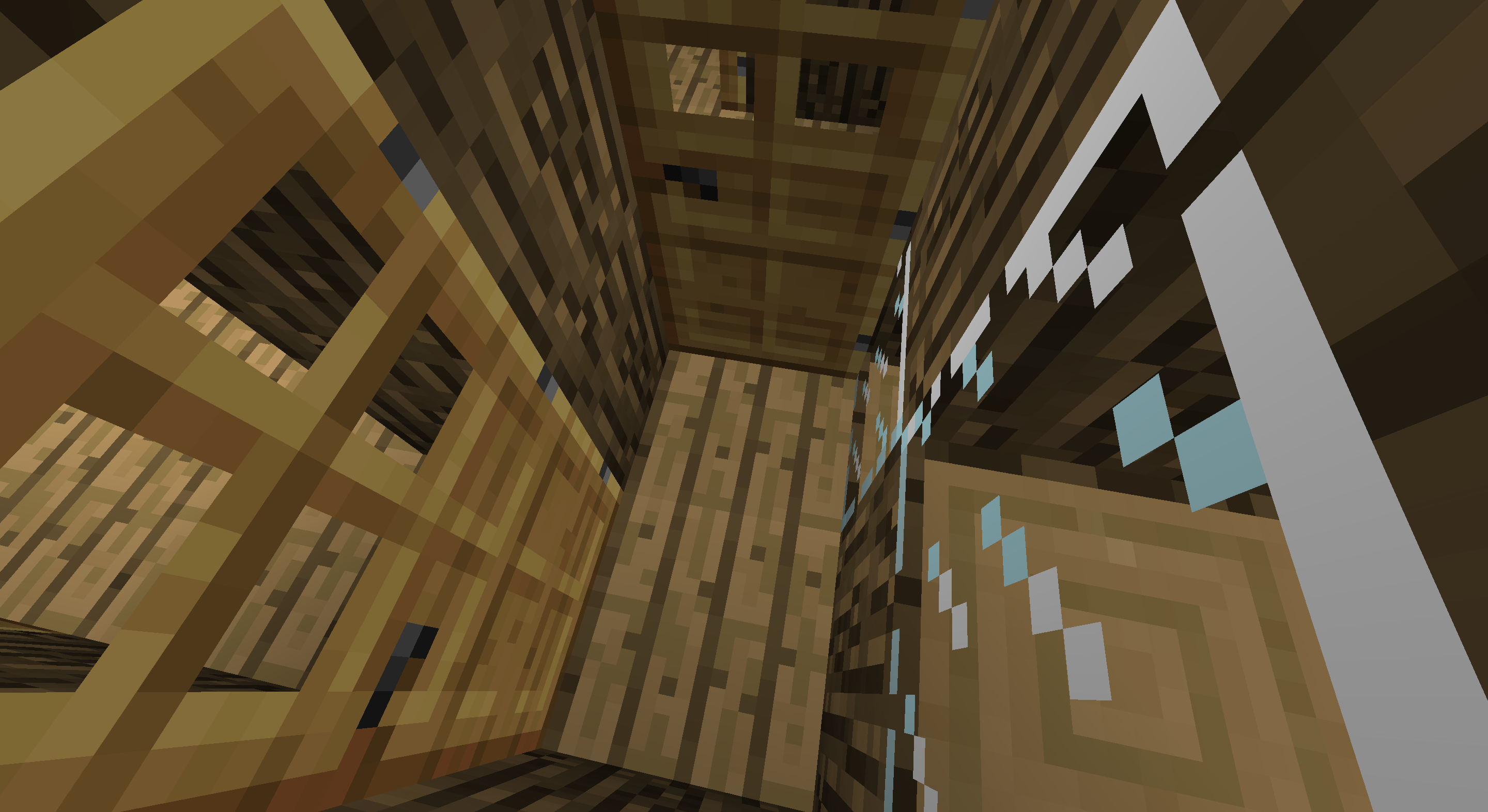}
    \caption{Inside a small room generated as a side effect of door placement}
    \label{fig:small-room2}
\end{figure}

An interesting side effect of the door placement algorithm is the creation of small ``closet rooms'' or ``pantries.'' Figures \ref{fig:small-room1} and \ref{fig:small-room2} display two such examples of these rooms. One could easily classify these rooms as pantries, closets, or even bathrooms.

The cellular-automata-generated external walls provide adequate amounts of natural light into the building, without forsaking privacy. Sometimes windows are placed very high up on the wall, while other times the windows stretch the entirety of the floor to the ceiling. Overall, this method provides dynamic and unique mosaics that are interesting to look at. Figure \ref{fig:living-room} displays one such example of diagonal wall patterns, and Figure \ref{fig:small-room2} displays a similar pattern within a ``small room.''

\section{Future Work}

One of the main aims of this work was to provide a reusable and extendable framework for building generation to the GDMC community. In the following section we want to discuss some possible extension that could be carried out, either by us or by others. 

\subsection{Modularity}

The generator executes several, modular stages, such a floor plan generation, door placement, and external wall generation. Each of these stages takes input from the previous step, but can be changed or extended in a modular way to create a greater variety of buildings, or buildings of a specific type. For example, one could provide something other than a rectangle for the initial building footprint. Or the expansion of rooms could be weighted, to create a range of smaller and bigger rooms. Similarly, the cellular automata that creates the walls can be modified to produce a different style of wall. It is even possible to just exchange one technique for another. So, instead of using a cellular automata, one could use a grammar based approach to create walls, while still using the other steps of the algorithm. 

\subsection{Adaptivity}

The adaptivity to existing content is a central challenge of the GDMC Settlement Generation competition. While we do not directly address it here we want to outline how the framework could be extended to tackle this. First, this approach is flexible enough to produce floorplans for arbitrary building footprints - and said footprints could be determined from available flat land on a given map. During later states it would also be possible to integrate existing terrain into the progress. For example, the cellular automate could also check the type of block immediately next to the house to determine if it is air or something else. This could then be used to not have windows show up next to an external dirt wall. The growth like approach can also grow around existing obstacles - imagine a multi story building that stands partially on support stilts. It might be a good idea to have those stilts extend thought the building in the form of pillars or load bearing walls. To realize this, we could set some piece of the floor plan to walls (possible even made from a sturdier material), and then still have the rooms grow around them. 


\section{Conclusion}
This paper introduces a simple yet effective way of floor-plan generation for Minecraft buildings. The method is a constrained growth approach, treating the rooms as individual entities which are allowed to grow one one block at a time. The generated buildings have an organic feel to them, differentiating themselves from traditional rectangular room layouts. As structures get larger, they often become disorienting due to the absence of any interior room landmarks. A cool side effect of the door placement algorithm produces ``small rooms,'' which are not originally designed in the constrained-growth-produced floor plan.

We hope to see this generator built on in future work. One of our motivations for this work was to push the current state of the GMDC settlement generator competition into moving away from template structure techniques and to use more procedurally generated ones. Our code is optimized to be used in place of a template, so we hope to see it or an augmented version of it used in a settlement generator. We would also like to improve this system by adding a room furnisher, which can add interior landmarks to help guide a player through the building. We believe this will help especially with the larger structures, which can sometimes be difficult to navigate.

\begin{acks}
Michael Cerny Green acknowledges the financial support of the SOE Fellowship from NYU Tandon School of Engineering. Christoph Salge is funded by the EU Horizon 2020 programme / Marie Sklodowska-Curie grant 705643
\end{acks}

%
\bibliographystyle{ACM-Reference-Format}
\bibliography{acmart}

\end{document}